\documentclass[twocolumn,10pt]{asme2e}

\usepackage{graphicx}
\usepackage{caption}

\title{Sketch2Prototype: Rapid Conceptual Design Exploration and Prototyping with Generative AI}

\author{Kristen M. Edwards\footnotemark[1]
    \affiliation{
    Massachusetts Institute of Technology\\
    Dept. of Mechanical Engineering\\
    kme@mit.edu
    }
}

\author{Brandon Man\footnotemark[1]
    \affiliation{ Massachusetts Institute of Technology\\
    Dept. of Mechanical Engineering\\
    bm557@mit.edu} }

\author{Faez Ahmed   
    \affiliation{    
    Massachusetts Institute of Technology\\
    Dept. of Mechanical Engineering\\
    faez@mit.edu
    }
}

\begin{document}



\twocolumn[{
\maketitle
\begin{center}
    \captionsetup{type=figure}
    \includegraphics[width=\textwidth]{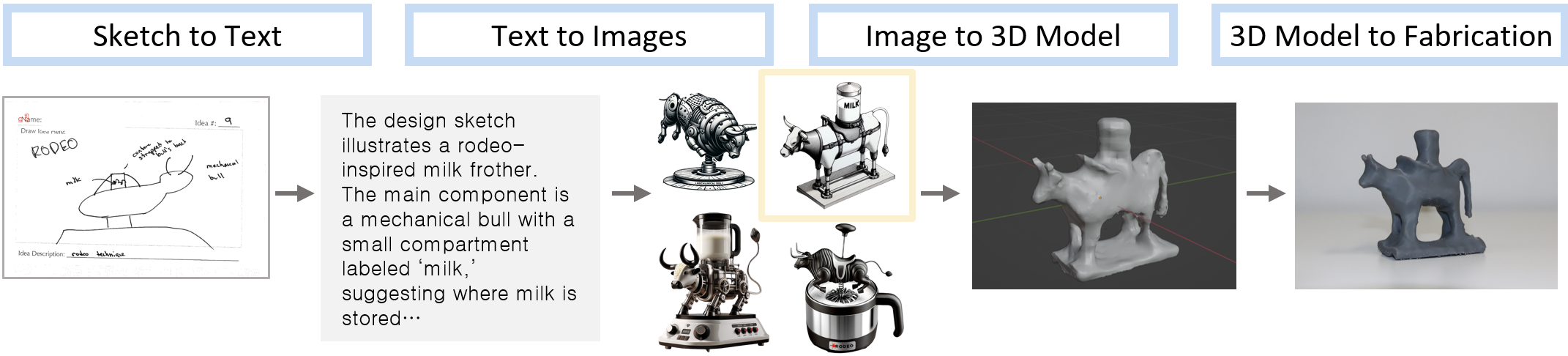}
    \captionof{figure}{The Sketch2Prototype framework takes in a conceptual design sketch as input, and produces multiple inspired images, a 3D model, and finally a fabricated prototype.}
\end{center}
}]


\footnotetext[1]{These authors contributed equally to this work.}

\begin{abstract}

Sketch2Prototype is an AI-based framework that transforms a hand-drawn sketch into a diverse set of 2D images and 3D prototypes through sketch-to-text, text-to-image, and image-to-3D stages. This framework, shown across various sketches, rapidly generates text, image, and 3D modalities for enhanced early-stage design exploration. We show that using text as an intermediate modality outperforms direct sketch-to-3D baselines for generating diverse and manufacturable 3D models. We find limitations in current image-to-3D techniques, while noting the value of the text modality for user-feedback and iterative design augmentation.

\end{abstract}


\section*{INTRODUCTION}
During product design and development, a design concept moves through many modalities. It may be represented as a sketch, a textual description, a looks-like or works-like prototype, and finally materialize as a finished product~\cite{ulrich2020product}. Early in the engineering design process, sketches and prototypes are pivotal for conveying ideas, investigating various design possibilities, and exploring the design space~\cite{bao2018interplay}. Developing a looks-like prototype is an essential step in the engineering design process, offering a tangible, visual representation of a product idea, and communicating the design concept to stakeholders. However, prototyping can be time-consuming and resource-intensive, involving multiple iterations and manual adjustments to achieve the desired outcome~\cite{lauff2019prototyping}.

 	In the phase of conceptual design, it's typical to progress from sketching to prototyping in a linear fashion. Yet, research indicates that tackling these activities concurrently can offer significant advantages~\cite{bao2018interplay}. Given that conceptual design determines up to 70-80\% of a product's lifetime cost~\cite{corbett1986design, pahl2007engineering} exploring the design space with proper breadth and depth is indeed valuable. Despite this, sketching and prototyping are often done in sequence because sketching is quicker and has lower overhead than prototyping~\cite{ulrich2020product}.

  Recent breakthroughs in generative AI have enabled people to generate novel, unseen images by learning underlying patterns in training data. Through generative models, the looks-like prototyping process can be streamlined, allowing for rapid generation and iteration of design options, thus significantly reducing time and costs associated with manual methods. This enables design space exploration, motivating designers with a diverse set of examples. Moreover, incorporating machine learning in the prototype development process opens the door for enhanced user interaction and feedback through easy iterations, as shown later in Figure \ref{fig:Pairwise CLIP}. Furthermore, communication between the development team can be bolstered by having a physical manifestation of their design vision. Ultimately, the integration of machine learning in creating looks-like prototypes facilitates a more efficient, cost-effective, and user-centered approach to product development, aligning technological innovation with aesthetic and practical design needs.

 	In this work, we propose Sketch2Prototype, a framework for understanding sketches, generating new conceptual images inspired by those sketches, converting the images to 3D models, and finally fabricating a looks-like prototype from these 3D models. There are various existing models that can perform each of these subtasks, we demonstrate several state-of-the-art methods. We found that GPT-4V(ision), a vision language model by OpenAI, 2023, is able to interpret and explain hand-drawn sketches~\cite{picard2023concept}. Therefore, we use GPT-4V, to convert sketches into textual prompt descriptions, then DALL-E 3, which is a generative text-to-image model, generates a set of more descriptive images from the text. We then use those images to generate a 3D model which, after postprocessing, we fabricated via additive manufacturing. We demonstrate Sketch2Prototype in a series of case studies with real hand drawn sketches of milk frothers, phone stands, pen and coin holders, and mugs. Our method enhances design space exploration by three means: 1) inherent design expansion caused from automatically generating multiple 2D images inspired by one sketch 2) increased breadth and depth of exploration made possible by working with sketches and prototypes in parallel~\cite{bao2018interplay}, and 3) allowing for user-centered feedback via the text modality. 
 	The contributions of this work are as follows:
\begin{enumerate}
    \item 	We introduced a generative AI-based framework to rapidly create a prototype from a sketch, enabling design exploration as the design moves through sketch, text, image, and 3D modalities.
    \item 	We compared our framework to direct sketch-to-3D and ControlNet-generated image-to-3D frameworks. Our model generates more diverse images and manufacturable designs.
    \item 	We demonstrated examples of the successful Sketch2Prototype, moving from a hand drawn sketch to a fabricated 3D looks-like prototype for four design categories and six designs.
    \item	We built an open-source dataset of 1,087 milk frother sketches each with four paired images inspired by the sketch and generated by our framework. This results in 4,348 images.
\end{enumerate}

\begin{figure*}
    \centering
    \includegraphics[width=\textwidth]{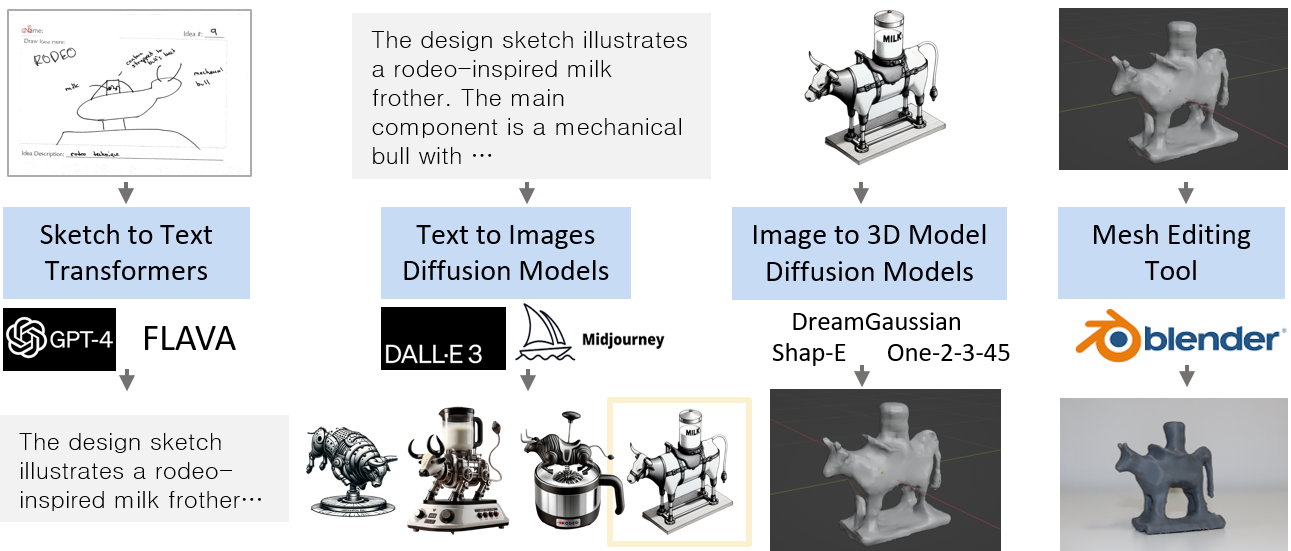}
    \caption{The Sketch2Prototype framework uses transformer-based models for the sketch-to- text and text-to-image steps, as well as an encoder and conditional diffusion model for image-
to-3D model. Post-processing of the 3D model is performed in Blender.}
    \label{fig:framework}
\end{figure*}

\section{RELATED WORK}
In the following sections we discuss related works regarding sketching and prototyping in engineering design, recent advancements in vision language models, image-to-3D models, and 3D representations. 

\subsection{Sketching and Prototyping in Engineering Design}

Sketching is documented as a valuable skill in engineering design, and researchers have studied ways to encourage and understand sketching in engineering education~\cite{schmidt2012research, das2022assessing}. Sketching provides a rapid external representation that comes at very little cognitive cost~\cite{goldschmidt2014modeling}. Researchers also explored creativity and decision making with sketches~\cite{toh2016choosing}, using sketching for finite element analysis~\cite{murugappan2017feasy}, and using machine learning to predict creativity-ratings from sketches and text~\cite{edwards2022picture, song2023attention}.

In product design, prototypes can be defined as ``an approximation of the product along one or more dimensions of interest''~\cite{ulrich2020product}. Prototypes, as well as the process of building and testing them, offer invaluable information to designers~\cite{das2022assessing}. As such, many works have surveyed and explored different prototyping strategies~\cite{camburn2017design, hansen2020from}. Research has explored how combining sketching and prototyping during conceptual design impacts design space exploration. On average, only sketching leads to a broader design space and generated more novel designs, only prototyping leads to more aesthetically pleasing designs with better functionally, both sketching and prototyping explored and generated final ideas that were perceived as more creative~\cite{bao2018interplay}. Sketching can lead to a higher quantity of designs~\cite{neeley2013building}, but prototyping can be used to both explore and refine designs~\cite{elverum2016prototyping}. Furthermore, this work suggests benefits of using both sketch and prototype modalities during conceptual design to explore the design space with both breadth and depth and ultimately generate creative designs.

One challenge that prevents designers from prototyping in parallel with sketching is that prototypes are often slower to create and have higher associated costs than sketching~\cite{lauff2019prototyping}, and designers are often reluctant to spend money and time on things when uncertainty is high, like early in the conceptual stage\cite{ulrich2020product}. This is where we believe the integration of machine learning to efficiently create looks-like prototypes presents a transformative opportunity.

\subsection{Large Language Models and Vision Language Models}
Large language models (LLMs) are billion-parameter transformers that are pre-trained on significant amounts of data, enabling them to perform a wide variety of natural language processing tasks such as translation, summarization and recognition. LLMs such as LLaMA \cite{touvron2023llama} learn to generate text aligned with human preferences through Reinforcement Learning with Human Feedback. 

Vision language models (VLMs) have also become quite popular. To create cohesive understanding between image and text, CLIP (Contrastive Language-Image Pre-training) \cite{radford2021learning} is a model that creates a joint embedding space between vision and language. CLIP is an efficient metric for measuring similarity between an image-text or image-image pair \cite{hessel2022clipscore}. Hence, VLMs such as GPT-4V and FLAVA \cite{singh2022flava} are similar to LLMs except that they train on multimodal datasets and often employ CLIP or similar methodologies learn a joint space between language and vision. VLMs need to understand complex relationships between text and image, thus requiring multimodal data for training. Although there is a certain degree of overlap in their applications, VLMs are distinctively advantageous for tasks necessitating visual comprehension, in contrast to LLMs, which excel in purely text-based endeavors. Given the inherently multimodal nature of early-stage design, VLMs are particularly well-suited for such applications.

Text-to-image synthesis has been explored via models like Imagen~\cite{saharia2022photorealistic}, DALL-E 3~\cite{dalle3}. Many researchers have used diffusion models for text-to-image tasks as diffusion models tend to be faster. Many text-to-image models also enable users control over their picture by prompting the model to change specific regions of an image via text, known as inpainting. However, inpainting for these models is often limited to text. 

To enable users to control image generation with their own sketches, models such as ControlNet and T2IAdapter have emerged. These models freeze the text-to-image model and use the users' sketches to guide the text-to-image generation process. However, these models give too much control to the users, and often results in images too similar to the users' sketch, meaning less exploration of the design space, as we show later, these models lead to fixation on the original concept. By leveraging VLMs' ability to provide text descriptions, which can then be expanded by text-to-image models, we generate a highly diverse set of prototypes.
 
\begin{figure*}[ht]
    \centering
    \includegraphics[width=\linewidth]{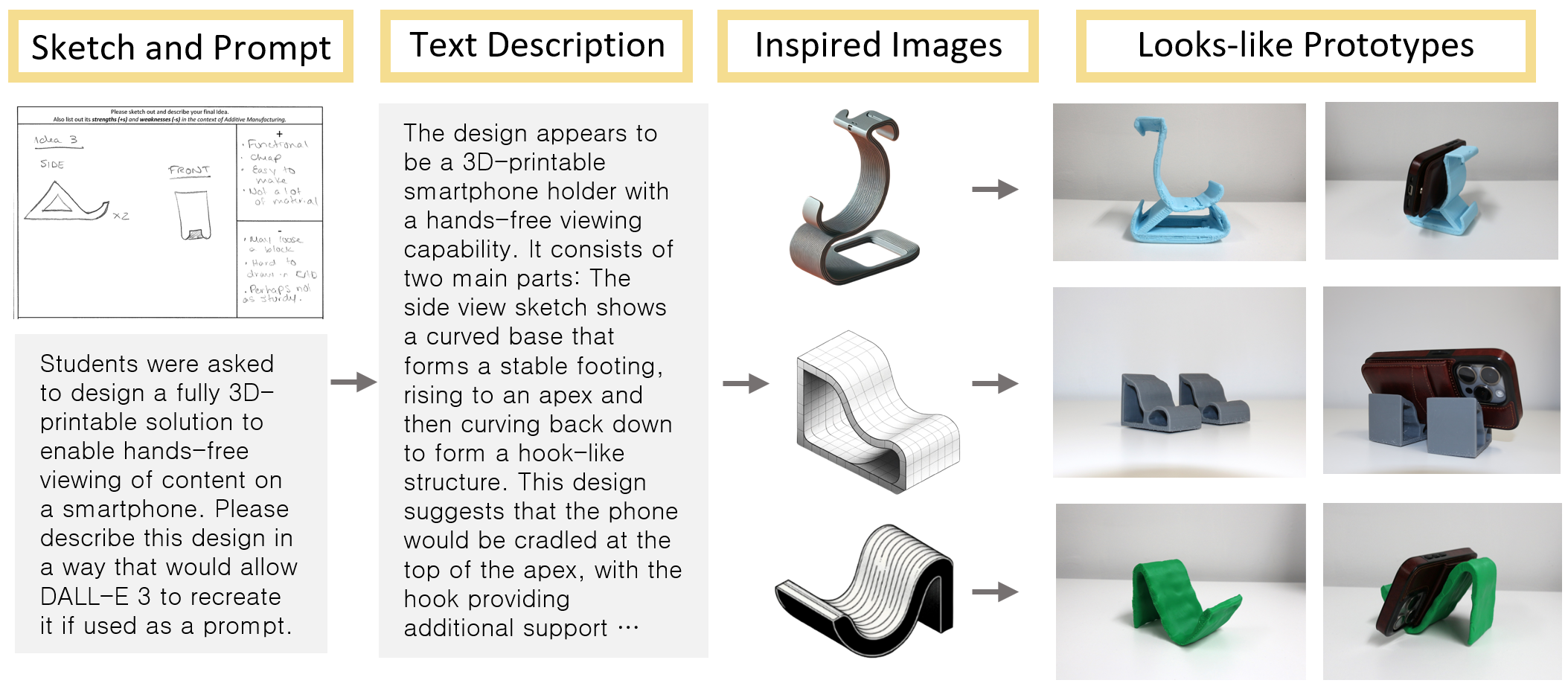}
    \caption{Our framework enables exploration of the design space by automatically generating multiple diverse images inspired by one sketch. Here a single sketch results in three fabricated looks-like prototypes.}
    \label{fig:phone_stand}
\end{figure*}

\subsection{3D Representations and Image-to-3D Models}

3D representations enable users to visualize the dimensions and proportions of objects. Furthermore, 3D representations enable manufacturing decisions, by displaying different geometric constraints. Lastly, 3D prototypes enable useful feedback from designers, customers, and stakeholders regarding the user-experience with a design \cite{das2022assessing}. Neural Radiance Fields (NeRFs) \cite{mildenhall2020nerf} have become a popular method for 3D scene representations. Although NeRFs are used in 3D reconstruction \cite{barron2021mipnerf} and generation \cite{poole2022dreamfusion} , optimizing NeRFs are time consuming to train and memory intensive. 3D Gaussian splatting \cite{kerbl3Dgaussians} is a recent alternative to NeRFs and has demonstrated promising results in both speed and quality in 3D reconstruction. Recent work has tried applying Gaussian Splatting to generation tasks \cite{tang2023dreamgaussian} that outperform methods that use NeRFs for 3D representations.

Image-to-3D models try to generate 3D assets from a single image, which can also be reformulated as a single-view 3D reconstruction tasks, but often produce blurry results \cite{yu2021pixelnerf}. Using image captioning models, text-to-3D methods can be adapted for image-to-3D generation \cite{melaskyriazi2023realfusion}. Dream Gaussian \cite{tang2023dreamgaussian} is a recent model that uses companioned mesh extraction and texture refinement in UV space. Even though the results of Dream Gaussian are promising, it fails to produce high quality models of unseen models. Shap-E \cite{jun2023shape} is another recent image-to-3D and text-to-3D model that utilizes a conditional diffusion model to output high fidelity 3D objects. 

In engineering design, it is often time consuming for designers to create high-quality images that can be used for image-to-3D generation tasks. Sketches are often abstract, lack detail, and are often unfit for image generation. Past work has tried to predict 3D functionality from a 2D image; however, 3D information performs best \cite{edwards2021design}. Our work shows an end-to-end system that generates multiple high-quality images from the original sketch, which can then generate a printable 3D prototype.

\section{METHODOLOGY}
In the following sections, we expound on the multi-stage process where a sketch is transformed into text, then to images. We further discuss the post-processing and fabrication stages, where 3D models are refined in Blender to meet fabrication standards and subsequently 3D printed to materialize the design concepts.

\subsection{Framework from Sketch to Prototype}
Our proposed framework treats the Sketch2Prototype problem as a sequence of tasks that move between design modalities: from sketch-to-text, then from text-to-image(s), and finally from image-to- 3D model, as shown in figure \ref{fig:framework}. For sketch-to-text, we fed our sketch as input along with a verbal description of the sketch into GPT-4V and prompted it to give it a description of the image. We also asked GPT-4V to describe it such that it will be passed as a prompt into DALL-E 3. DALL-E 3 performs text-to-image by converting the text description of the original sketch into a text embedding, then feeding it into a diffusion prior to generate an image embedding, which finally gets decoded into an image. We chose not to extract the words found on the sketches as they may give semantic meaning to specific areas of the image. To generate a variety of novel designs, we ask DALL-E 3 to generate 4 images from the original prompt. DALL-E 3 also attempts to generate more diverse images by rephrasing the input prompt.

\begin{figure*}
    \centering
    \includegraphics[width = \linewidth]{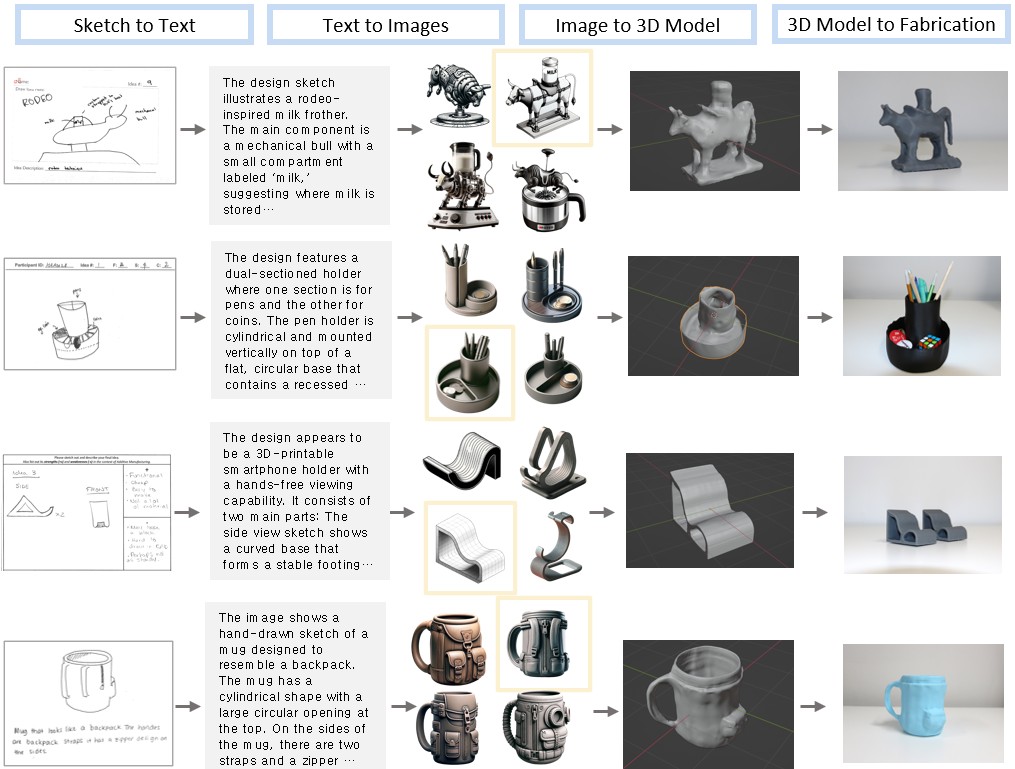}
    \caption{Examples of the full Sketch2Prototype framework for four different design types: rodeo-inspired milk frother, pen-and-coin holder, phone stand, and backpack-inspired mug.}
    \label{fig:fullExample}
\end{figure*}

The resulting image generated from DALL-E 3 may contain text, which negatively affects the generation quality when converting from image to 3D. To prevent this, we manually select a set of images that do not include any text and feed it into our image-to-3D model. Current state-of-the-art models such as Shap-E \cite{jun2023shape} and DreamGaussian have varied performance depending on the provided image. As a result, we feed our images into three state-of-the-art models (One-2-3-45, DreamGaussian and Shap-E) and pick the mesh that is most similar to the original image while also being the most manufacturable.

\subsection{Post-processing and Fabrication}
While models like Shap-E excel at generating 3D models, they may not adhere to fabrication requirements. Hence, we perform post-processing of the 3D model in Blender 3.6.5. Shap-E outputs a 3D model in the Polygon File Format (PLY) family. This can be directly imported to Blender 3.6.5, post-processed as needed, and exported as an STL file, which can be 3D printed. We print the models with either the Bambu Lab X1-Carbon Combo 3D Printer, which is a fused deposition modelling printer, or Formlabs Form 3 Printer, which is a stereolithography printer.


\begin{figure*}
    \centering
    \includegraphics[width = \linewidth]{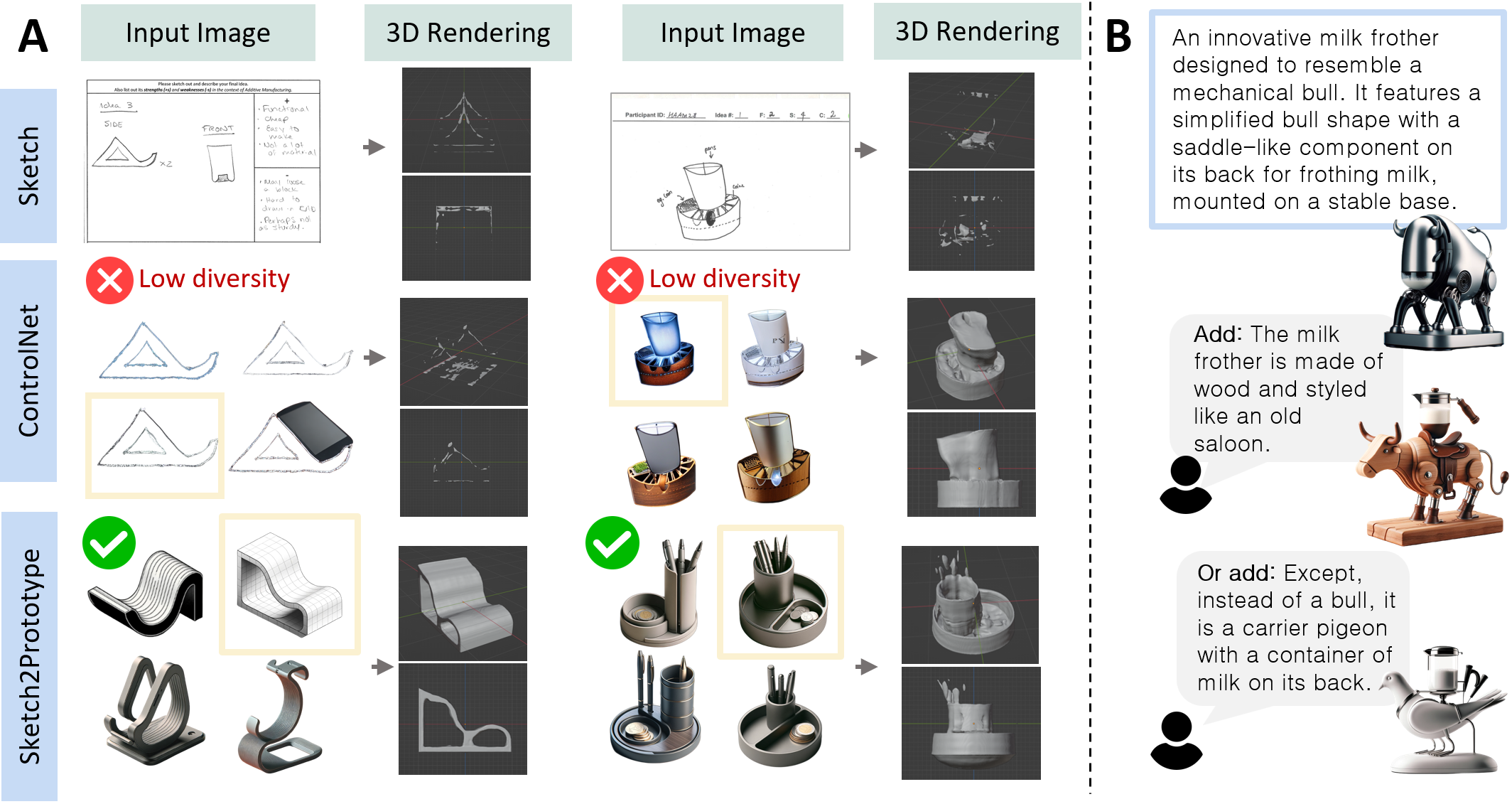}
    \caption{\textbf{A:} 3D models generated from varying input images: Sketch2Prototype generates more diverse and manufacturable designs. \textbf{B:} The text modality allows for user control. We append
text to the original prompt to generate different designs.}
    \label{fig:comparison}
\end{figure*}
\section{RESULTS}
In this section, we display the results of our framework via a number of examples. Figure \ref{fig:phone_stand} demonstrates how the Sketch2Prototype framework enables exploration of the design space. From one design sketch of a phone stand, our framework leads to three diverse prototypes. The images and 3D printed models show a diverse set of functional phone stands. Figure \ref{fig:fullExample} showcases the full Sketch2Prototype framework for four different design types: a rodeo-inspired milk frother, a pen-and- coin holder, a phone stand, and a backpack-inspired mug. For each of these, we demonstrate an automatic exploration of the design space in the text-to-image step. Here, using a text description created from a sketch via generative-AI, a designer is automatically presented with any number of detailed design images inspired by the sketch. We chose to display four images for each sketch, however there is no imposed limit on this. A benefit of design exploration in this stage is that it mitigates design fixation, and can thus be used as an assistant for designers. Furthermore, showing a designer multiple diverse examples can aid in creative ideation. To assess the diversity and feasibility of our designs, we compare the diversity and manufacturability of our model's designs to those made from a sketch alone or using ControlNet, which adheres to the sketch geometry (Figure \ref{fig:comparison}A).

\subsection{Enhanced diversity and manufacturability over baselines}
We perform a qualitative evaluation of Sketch2Prototype by generating 3D models with two baseline approaches. The first approach is directly passing an unprocessed sketch into Shap-E to generate the resulting mesh. The second approach is passing our sketch into ControlNet to generate 4 candidate images. We then pass each image into Shap-E and, for standardization, select the first generated mesh. For Sketch2Prototype, we also generate 4 candidate images and perform the same mesh selection process. We test our method on a phone stand design and a pen-and-coin holder. Sketch-to-text via GPT-4V, text-to-image via DALL-E 3 or ControlNet, and image-to-3D via Shap-E each take a matter of seconds, so the time difference between these three approaches is negligible.

Results are shown in Figure \ref{fig:comparison}A. We can see that the meshes generated from unprocessed sketches are sparse and unmanufacturable. For ControlNet, the generated images lack diversity. In the case of the phone stand, due to the simplicity of the sketch, this process also produces unmanufacturable designs. The ControlNet generated images lack diversity due to directly matching the sketch geometry. Finally, our model generates diverse variations of each sketch. Our model also generated functional and cohesive meshes for prototyping.

\subsection{Generation and exploration via human-in-the-loop feedback}

We also evaluate the controllability of Sketch2Prototype via the text modality. In Figure \ref{fig:comparison}B, we add sentences to the DALL-E 3 prompt to alter the output image according to designer feedback. For example, when the original prompt is appended with The milk frother is made of wood and styled like and old saloon,'' the output image changes form accordingly. The intermediate text modality thus enables users to add iterative feedback, allowing for extra user control. Hence, there is immense value in using text as an intermediary modality to help with exploration and addition of new requirements, which are difficult to update in sketch directly.
\begin{figure}
    \centering
    \includegraphics[width=\linewidth]{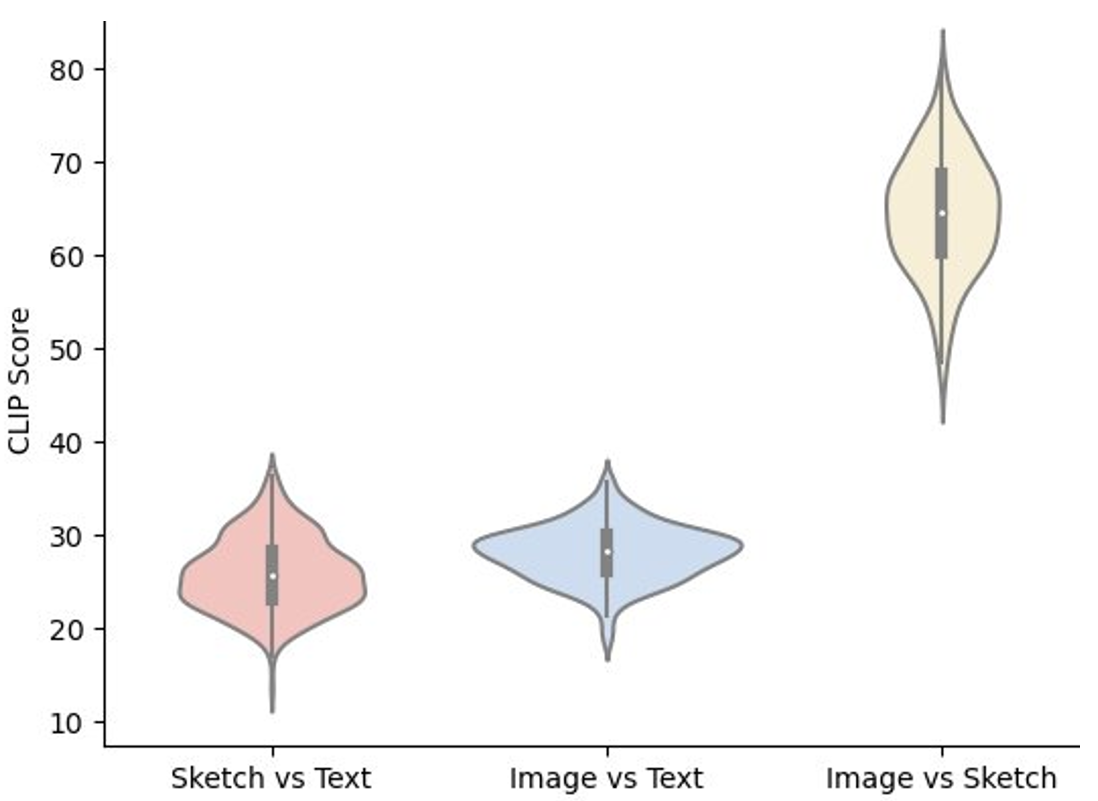}
    \caption{CLIP scores between Sketch and Text, Image and Text, and Image and Sketch modalities for sets of the sketch, text, and images corresponding to the same design. A high CLIP score indicates a high level of similarity and alignment between the respective pairs.}
    \label{fig:CLIP}
\end{figure}

\begin{figure}
    \centering
    \includegraphics[width=\linewidth]{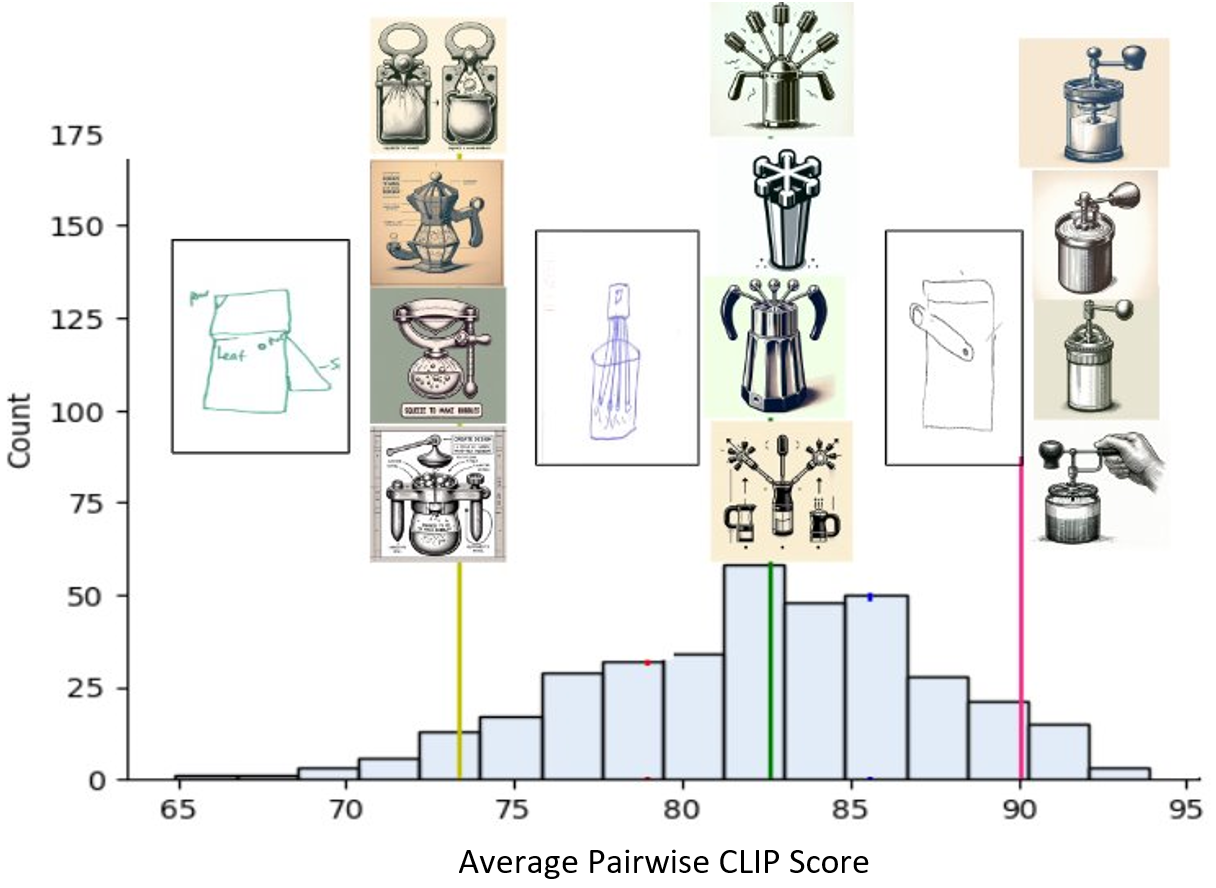}
    \caption{Average pairwise CLIP score between the four generated images. A lower CLIP score indicates a more diverse set of images. Yellow, red, green, blue and pink lines correspond to 5th, 50th, and 95th percentiles respectively. The original sketches are shown to the left of the set of images generated from them.}
    \label{fig:Pairwise CLIP}
\end{figure}

\subsection{Dataset alignment and diversity}
We show quantitatively that Sketch2Prototype generates images that match the original description while being diverse. We generated a synthetic dataset from a set of milk frother designs and evaluated its alignment with the original dataset and its diversity. This dataset of milk frothers contains 359 hand-drawn sketches of milk frothers drawn by unique individuals. Each milk frother design often has a brief text description of the design concept in addition to text annotations on the sketch. To generate the synthetic dataset, we provided the image to GPT-4V and incorporated the text description into prompt that asks GPT-4V to generate a DALL-E 3 prompt. For each sketch, we generate 4 images, giving us a synthetic dataset of 1,436 total images.

For our generated dataset, we show 1) that the generated images align with their sketches, and 2) that the generated images represent a diverse set, thus exploring the design space. To show our synthetic dataset aligns with the original dataset, we compute the CLIP score between each sketch in the original dataset and the provided text description. We also take the average CLIP score between the 4 images in the augmented dataset with the text description. Finally, we take the CLIP score between the sketches and the images. The results are listed in the plot on Figure \ref{fig:CLIP}.

The resulting average CLIP score for the sketch vs text, image vs text and sketch vs image sets are 25.8, 28.1 and 64.4 respectively. The higher average between the image vs text and sketch vs text is expected, since GPT-4V gave a more detailed text representation for generation compared to the original sketch. Likewise, CLIP scores between two images are generally higher than CLIP scores between image and text, so it is expected that the image-sketch CLIP score is higher than the remaining two.

To measure diversity, we compute the average pairwise CLIP score for each of the 4 images in each sample. Higher CLIP scores indicate less diverse datasets. Figure \ref{fig:Pairwise CLIP} illustrates the distribution of these pairwise CLIP scores. At the 5th, 50th, and 95th percentiles, we show an example of a sketch and four generated images to give better context on how these scores correspond to the diversity images. As percentile increases, we observe a decrease in geometric variation and fewer components. At the 95th percentile, the milk frother designs shown have almost identical geometries - every design are variations of a cylindrical container attached to a handle. However, designs at the 5th percentile have varied shapes and multiple components, such as handles of different curvatures, and different containers.


\section{DISCUSSION}
\subsection{Integrating AI into the Design Thinking Process}
In this work, we proposed a framework, Sketch2Prototype, that utilizes generative AI to rapidly generate a textual description, a diverse set of 2D images, and 3D models from a hand drawn sketch. We compare our framework to two baseline models, using sketch-alone and using ControlNet-generated images. We showed that our framework generates more diverse designs and manufacturable models than the others (Figure \ref{fig:comparison}A). Sketch2Prototype also increases the breadth and depth of exploration by allowing designers to work with sketches and prototypes in parallel. We exhibit the entire framework, resulting in six fabricated prototypes from four hand-drawn sketches.

We also demonstrate how user feedback can be incorporated via the text modality. Our framework represents a design as a sketch, text, image, and 3D model. The intermediate text modality can be easily edited by a designer to add more requirements that are not present in original text. This allows iterative refinement and improvement, shown in Figure \ref{fig:comparison}B. The two results shown in Figure \ref{fig:comparison} demonstrate the balance between user-control and automatic design expansion. While ControlNet allows for strict geometric adherence to an input image, this eliminates most diversity and may not be desirable when dealing with an imprecise hand-drawn sketch. On the other hand, DALL-E 3 generates very diverse designs, however iterative feedback via text may be necessary to generate a desired image.

Tools such as Sketch2Prototype can be incorporated into traditional design thinking processes, enhancing ideation, and prototype development. Sketch2Prototype enables engineers to rapidly explore the design space by expanding simple, abstract sketches into diverse images and 3D printable looks-like prototypes. Our framework allows for human-in-the-loop feedback in the text-to-image phase, and facilitates prototype development by converting these images into 3D models.

\subsection{Limitations within the existing framework}
We must emphasize that this is a preliminary exploration of how emerging AI tools may benefit designers via accelerated design transformation from sketch to text, images, and 3D models. The 3D models are meant to be looks-like, not functional, prototypes that can be built using additive manufacturing techniques. Further studies are necessary to determine which AI tools perform best at each step of the framework, and, in fact, define what ``best'' means.

The process of combining different design transformation steps means that deviance from the desired design can be introduced at every step. In the sketch-to-text phase, the user benefits from having much control via simple text editing. However, we observe a failure case in sketch-to-text then text-to- image when GPT-4V generates a text description that is deemed an ``unsafe'' prompt by DALL-E 3. This indicates that GPT-4V's text generation is not grounded on DALL-E 3's safety mechanisms.

Another limitation is the lack of repeatability in the text-to-image stage. Using off-the-shelf image generation tools such as DALL-E 3 means that the same prompt will not generate the same image when repeated. Furthermore, changes to the design are often global rather than local, even if the textual prompt only requests a local change.

The greatest limitation that we observe is in the image-to-3D stage. Though we utilize state of the art image-to-3D models, design details are often lost at this stage, and 3D models generated using image-to-3D are often non-smooth, fragmented, and sometimes non-manufacturable. As a result, postprocessing is required either to smooth surfaces, fill holes, or remove unmanufacturable parts. Even with postprocessing, image-to-3D models tend to create uneven surfaces or holes, which indicate that NeRFs, while strong for 3D visualization purposes, may not be ideal for 3D printing. Due to the postprocessing step, the time taken for Sketch2Prototype increases.

Finally, aside from manufacturability problems, there may be a lack of control over the final mesh. For earlier steps in the framework, users can edit intermediate representations to better control the product, such as editing descriptions during sketch-to-text or inpainting during text-to-image. The only control users have over a mesh is the postprocessing stage, which is limited to removal or minor edits to surfaces. Enabling users to ``inpaint'' 3D meshes would give significantly more flexibility over their final product.

\section{FUTURE WORK}

Future research could explore the application of this framework in more complex design scenarios, such as multi-component systems or intricate structures. These tasks are challenging since models would need to identify distinct parts, understand part-interfaces, and ensure compatibility. This challenge becomes even harder when dealing with sketches that have internal components, as internal components need to have correct proportions with respect to their container in order to fit, and we note that the image-to-3D models we use did not capture internal components well.

Current research on image-to-3D models is often concerned with synthesizing 3D images from objects for visualization purposes, but making these meshes functional is much harder. This work reveals that NeRFs may not be an ideal candidate for representing manufacturable prototypes. An area for exploration may be to create a new representation of meshes specifically for 3D printing purposes.

This work aims to demonstrate how existing AI tools enable transformation from sketch to text, image, and 3D modalities, which can enhance design space exploration. Sketch2Prototype should assist designers in efficiently ideating with different modalities. To this end, future work may include incorporating user-centered design principles in the Sketch2Prototype framework, focusing on intuitive interfaces and user feedback.

\section{CONCLUSION}
We demonstrate a framework to convert sketches into fabricated prototypes via intermediate steps: sketch-to-text, text-to-image, and image-to-3D. We show that our framework enhances design space exploration by generating a set of diverse 2D images and 3D models from a single sketch, and by enabling designers to work with sketches and prototypes in parallel. We find that using text as an intermediate modality allows for iterative user feedback and enhanced user control. Furthermore, text- to-image-to-3D generates more diverse and manufacturable 3D models than sketch-to-3D baselines. However, manufacturability is still a limitation of current image-to-3D models. The Sketch2Prototype framework gains potential as each step is actively worked on by the machine learning community.

\bibliographystyle{asmems4}

\begin{acknowledgment}
This material is based upon work supported by the National Science Foundation under Grant No. 2231254 and the NSF Graduate Research Fellowship.
\end{acknowledgment}

%
% ...

\bibliography{asme2e}

\begin{thebibliography}{10}

\bibitem{ulrich2020product}
Ulrich, K.~T., Eppinger, S.~D., and Yang, M.~C., 2020.
\newblock {\em Product Design and Development}.
\newblock McGraw-Hill Education, New York, NY.

\bibitem{bao2018interplay}
Bao, Q., Faas, D., and Yang, M., 2018.
\newblock ``Interplay of sketching \& prototyping in early stage product design''.
\newblock {\em International Journal of Design Creativity and Innovation, \textbf{ 6}}(3-4), pp.~146--168.

\bibitem{lauff2019prototyping}
Lauff, C., Menold, J., and Wood, K., 2019.
\newblock ``Prototyping canvas: Design tool for planning purposeful prototypes''.
\newblock In Proceedings of the Design Society: International Conference on Engineering Design, Vol.~1, pp.~1563--1572.

\bibitem{corbett1986design}
Corbett, J., and Crookall, J., 1986.
\newblock ``Design for economic manufacture''.
\newblock {\em CIRP Annals, \textbf{ 35}}(1), pp.~93--97.

\bibitem{pahl2007engineering}
Pahl, G., Beitz, W., Feldhusen, J., and Grote, K.-H., 2007.
\newblock {\em Engineering Design: A Systematic Approach}.
\newblock Springer London.

\bibitem{picard2023concept}
Picard, C., Edwards, K., Doris, A., Man, B., Giannone, G., Alam, M., and Ahmed, F., 2023.
\newblock ``From concept to manufacturing: Evaluating vision-language models for engineering design''.
\newblock {\em arXiv preprint arXiv:2311.12668}.

\bibitem{schmidt2012research}
Schmidt, L., Hernandez, N., and Ruocco, A., 2012.
\newblock ``Research on encouraging sketching in engineering design''.
\newblock {\em AI EDAM (Artificial Intelligence for Engineering Design, Analysis and Manufacturing), \textbf{ 26}}(3), pp.~303--315.

\bibitem{das2022assessing}
Das, M., and Yang, M., 2022.
\newblock ``Assessing early stage design sketches and reflections on prototyping''.
\newblock {\em ASME Journal of Mechanical Design, \textbf{ 144}}(4), p.~041403.

\bibitem{goldschmidt2014modeling}
Goldschmidt, G., 2014.
\newblock ``Modeling the role of sketching in design idea generation''.
\newblock In {\em Anthology of Theory and Models of Design}, A.~Chakrabarti and L.~Blessing, eds. Springer, London.

\bibitem{toh2016choosing}
Toh, C., and Miller, S., 2016.
\newblock ``Choosing creativity: the role of individual risk and ambiguity aversion on creative concept selection in engineering design''.
\newblock {\em Research in Engineering Design, \textbf{ 27}}, pp.~195--219.

\bibitem{murugappan2017feasy}
Murugappan, S., Piya, C., Yang, M., and Ramani, K., 2017.
\newblock ``Feasy: A sketch-based tool for finite element analysis''.
\newblock {\em ASME Journal of Computing and Information Science in Engineering, \textbf{ 17}}(3), p.~031009.

\bibitem{edwards2022picture}
Edwards, K., Peng, A., Miller, S., and Ahmed, F., 2022.
\newblock ``If a picture is worth 1000 words, is a word worth 1000 features for design metric estimation?''.
\newblock {\em ASME Journal of Mechanical Design, \textbf{ 144}}(4), p.~041402.

\bibitem{song2023attention}
Song, B., Miller, S., and Ahmed, F., 2023.
\newblock ``Attention-enhanced multimodal learning for conceptual design evaluations''.
\newblock {\em ASME Journal of Mechanical Design, \textbf{ 145}}(4), p.~041410.

\bibitem{camburn2017design}
Camburn, B., Viswanathan, V., Linsey, J., Anderson, D., Jensen, D., Crawford, R., Otto, K., and Wood, K., 2017.
\newblock ``Design prototyping methods: State of the art in strategies, techniques, and guidelines''.
\newblock {\em Design Science, \textbf{ 3}}, p.~E13.

\bibitem{hansen2020from}
Hansen, C., and Özkil, A., 2020.
\newblock ``From idea to production: A retrospective and longitudinal case study of prototypes and prototyping strategies''.
\newblock {\em ASME Journal of Mechanical Design, \textbf{ 142}}(3), p.~031115.

\bibitem{neeley2013building}
Neeley, L., Lim, K., Zhu, A., and Yang, M., 2013.
\newblock ``Building fast to think faster: Exploiting rapid prototyping to accelerate ideation during early stage design''.
\newblock In ASME International Design Engineering Technical Conferences.

\bibitem{elverum2016prototyping}
Elverum, C., Welo, T., and Tronvoll, S., 2016.
\newblock ``Prototyping in new product development: Strategy considerations''.
\newblock {\em Procedia CIRP, \textbf{ 50}}, pp.~117--122.

\bibitem{touvron2023llama}
Touvron, H., Lavril, T., Izacard, G., Martinet, X., Lachaux, M.-A., Lacroix, T., Rozière, B., Goyal, N., Hambro, E., Azhar, F., Rodriguez, A., Joulin, A., Grave, E., and Lample, G., 2023.
\newblock Llama: Open and efficient foundation language models.

\bibitem{radford2021learning}
Radford, A., Kim, J.~W., Hallacy, C., Ramesh, A., Goh, G., Agarwal, S., Sastry, G., Askell, A., Mishkin, P., Clark, J., Krueger, G., and Sutskever, I., 2021.
\newblock Learning transferable visual models from natural language supervision.

\bibitem{hessel2022clipscore}
Hessel, J., Holtzman, A., Forbes, M., Bras, R.~L., and Choi, Y., 2022.
\newblock Clipscore: A reference-free evaluation metric for image captioning.

\bibitem{singh2022flava}
Singh, A., Hu, R., Goswami, V., Couairon, G., Galuba, W., Rohrbach, M., and Kiela, D., 2022.
\newblock Flava: A foundational language and vision alignment model.

\bibitem{saharia2022photorealistic}
Saharia, C., Chan, W., Saxena, S., Li, L., Whang, J., Denton, E., Ghasemipour, S. K.~S., Ayan, B.~K., Mahdavi, S.~S., Lopes, R.~G., Salimans, T., Ho, J., Fleet, D.~J., and Norouzi, M., 2022.
\newblock Photorealistic text-to-image diffusion models with deep language understanding.

\bibitem{dalle3}
Betker, J., Goh, G., Jing, L., Brooks, T., Wang, J., Li, L., Ouyang, L., Zhuang, J., Lee, J., Guo, Y., Manassra, W., Dhariwal, P., Chu, C., Jiao, Y., and Ramesh, A., 2022.
\newblock Improving image generation with better captions.

\bibitem{mildenhall2020nerf}
Mildenhall, B., Srinivasan, P.~P., Tancik, M., Barron, J.~T., Ramamoorthi, R., and Ng, R., 2020.
\newblock Nerf: Representing scenes as neural radiance fields for view synthesis.

\bibitem{barron2021mipnerf}
Barron, J.~T., Mildenhall, B., Tancik, M., Hedman, P., Martin-Brualla, R., and Srinivasan, P.~P., 2021.
\newblock Mip-nerf: A multiscale representation for anti-aliasing neural radiance fields.

\bibitem{poole2022dreamfusion}
Poole, B., Jain, A., Barron, J.~T., and Mildenhall, B., 2022.
\newblock Dreamfusion: Text-to-3d using 2d diffusion.

\bibitem{kerbl3Dgaussians}
Kerbl, B., Kopanas, G., Leimk{\"u}hler, T., and Drettakis, G., 2023.
\newblock ``3d gaussian splatting for real-time radiance field rendering''.
\newblock {\em ACM Transactions on Graphics, \textbf{ 42}}(4), July.

\bibitem{tang2023dreamgaussian}
Tang, J., Ren, J., Zhou, H., Liu, Z., and Zeng, G., 2023.
\newblock ``Dreamgaussian: Generative gaussian splatting for efficient 3d content creation''.
\newblock {\em arXiv preprint arXiv:2309.16653}.

\bibitem{yu2021pixelnerf}
Yu, A., Ye, V., Tancik, M., and Kanazawa, A., 2021.
\newblock pixelnerf: Neural radiance fields from one or few images.

\bibitem{melaskyriazi2023realfusion}
Melas-Kyriazi, L., Rupprecht, C., Laina, I., and Vedaldi, A., 2023.
\newblock Realfusion: $360{\deg}$ reconstruction of any object from a single image.

\bibitem{jun2023shape}
Jun, H., and Nichol, A., 2023.
\newblock Shap-e: Generating conditional 3d implicit functions.

\bibitem{edwards2021design}
Edwards, K., Addala, V., and Ahmed, F., 2021.
\newblock ``Design form and function prediction from a single image''.
\newblock In ASME International Design Engineering Technical Conferences and Computers and Information in Engineering Conference.

\end{thebibliography}



\end{document}